# A Novel Type of Very Long Baseline Astronomical Intensity Interferometer

Ermanno F. Borra,
Centre d'Optique, Photonique et Laser,
Département de Physique, Université Laval, Québec, Qc, Canada G1K 7P4
(email: borra@phy.ulaval.ca)

**KEY WORDS** : Techniques: interferometric

# ABSTRACT

This article presents a novel type of very long baseline astronomical interferometer that uses the fluctuations, as a function of time, of the intensity measured by a quadratic detector, which is a common type of astronomical detector. The theory on which the technique is based is validated by laboratory experiments. Its outstanding principal advantages comes from the fact that the angular structure of an astronomical object is simply determined from the visibility of the minima of the spectrum of the intensity fluctuations measured by the detector, as a function of the frequency of the fluctuations, while keeping the spacing between mirrors constant. This would allow a simple setup capable of high angular resolutions because it could use an extremely large baseline. Another major interest is that it allows for a more efficient use of telescope time because observations at a single baseline are sufficient, while amplitude and intensity interferometers need several observations at different baselines. The fact that one does not have to move the telescopes would also allow detecting faster time variations because having to move the telescopes sets a lower limit to the time variations that can be detected. The technique uses wave interaction effects and thus has some characteristics in common with intensity interferometry. A disadvantage of the technique, like in intensity interferometry, is that it needs strong sources if observing at high frequencies (e.g. the visible). This is a minor disadvantage in the radio region. At high frequencies, this disadvantage is mitigated by the fact that, like in intensity interferometry, the requirements of the optical quality of the mirrors used are far less severe than in amplitude interferometry so that poor quality large reflectors (e.g. Cherenkov telescopes) can be used in the optical region.

# 1. INTRODUCTION

Most interferometry used in astronomy is based on techniques that use first order correlations. We will refer to it as amplitude interferometry. Hanbury Brown and Twiss (Hanbury Brown 1968) introduced intensity interferometry, a technique based on second order correlations.

While amplitude and intensity interferometry are described in textbooks (e.g. Klein & Furtak 1986), it is less known that interfering beams that have large optical path differences, and therefore give an interferometric signal unobservable with amplitude and intensity interferometers, give a recombined beam possessing a spectral distribution modulated by periodic minima and maxima. Spectral modulation occurs for optical path differences far larger than the coherence length of the interfering beams. This surprising statement is supported by experiments and theoretical analyses. Alford and Gold (1958) used a visible light source and found a periodic modulation of the spectrum of current fluctuations measured with a photomultiplier for an optical path difference of about 30 meters, far exceeding the coherence length of the source (an electric spark). Mandel (1962) gives a full theoretical justification, while Givens (1961) gives a less rigorous but easier to follow physical explanation. While Alford and Gold (1958) used pulsed sources, Givens (1961) predicted that the spectral modulation should also be present for interfering beams from continuous sources. This was experimentally confirmed by Basano & Ottonello (2000).

In this article, I propose a novel type of astronomical interferometer that measures the visibility of the periodic modulation of the spectrum of the fluctuations of the output current (intensity) of a quadratic detector that measures the combined beams from separate telescopes. I also propose to use numerical analyses of separate digital recording of the output signals from distant telescopes: This should allow extremely large baselines. Note that a quadratic detector is a detector that measures the time average of the square of the electric field. It is therefore a very common type of astronomical detector (e.g. a photomultiplier). The theoretical foundations of the interferometric technique are contained in Borra (2008) and are experimentally confirmed by Borra (2011). Note that Borra (2008) models a gravitational lens by a Young interferometer (a standard model for gravitational lenses), therefore his theoretical analysis is obviously valid in the context of astronomical interferometry.

Like with intensity interferometry, the theoretical basis of the technique may be difficult to understand for people unfamiliar with statistical optics. The simple model of a disk having a constant angular intensity distribution observed by the interferometer discussed in section 2 below gives an intuitive understanding.

## 2. THE INTERFEROMETER

This section presents the interferometer and discusses techniques that can be used to obtain the spectrum of the intensity fluctuations.

The physical basis of the interferometer can be found in Borra (2008) who discusses an observational technique to obtain time delays in gravitational lenses. Borra (2008) models gravitational lenses with a Young interferometer, which is a standard model for gravitational lenses: Consequently his theoretical analysis also applies to astronomical interferometers. Figure 1 in Borra (2008) shows the Young interferometer model, where the impact parameter *a* is equivalent to the separation between the slits of a Young interferometer. The recombined beams are observed by a telescope and measured by a quadratic detector, which is a detector that measures the time average of the electric field. The output of the quadratic detector is the intensity as a function of time *I(t)*. This type of detector is routinely used in astronomical telescopes. For example, photomultipliers or CCD detectors are quadratic detectors. Finally a wave analyzer obtains the frequency spectrum *I($\omega$')* of the output current *I(t)* measured by quadratic detector. The beat frequency *$\omega$'* is the frequency of the spectrum of the fluctuations of the output current of the quadratic detector, is produced by waves in the spectrum of the source beating among themselves, and is a much lower frequency than the frequency of observation *$\omega$* in the frequency spectrum of the source. The visibility of the spectral modulation of *I($\omega$')* depends on the angular intensity distribution of the source *i($\theta$)* and, consequently, can be used to find it

In an amplitude interferometer the normalized correlation function quantifies the visibility of the intensity fringes. The visibility of the spectral modulation *I($\omega$')* is also quantified by an analogous normalized correlation function that can also be written as the product *$\gamma$(0)$\gamma$($\tau$)* of the two terms commonly called spatial and temporal coherence functions. Like in amplitude interferometry, the spatial coherence function *$\gamma$(0)* is given by the Fourier transform of the normalized angular intensity distribution of the source *i($\theta$)* and must be evaluated at a specific frequency. However, in our case we have two different frequencies: *$\omega$* the frequency of the spectrum measured by the detector (e.g. *$10^{14}$ Hz*) and *$\omega$'*, the beat frequency of the spectrum of the current fluctuations of the output current of the detector (e.g. *$10^{10}$ Hz*). In amplitude interferometry the spatial coherence function is evaluated at the frequency measured by the detector. Borra (1997) made this assumption when first suggesting the technique. However Borra (2008) reexamined the problem from first principles, starting from the superposition of electromagnetic waves, and demonstrated that, in our case, *$\gamma$(0)* must be evaluated at the *$\omega$'* frequency. Borra (2011) carried out experiments that validate the theory in Borra (2008).

The theoretical analysis in Borra (2008) may be difficult to understand for people unfamiliar with statistical optics. The discussion leading to equation 16 in Borra (2008), summarized below, gives an intuitive feeling for the basis of the technique. Borra (2008) used a simple source which is a disk having angular diameter *$\Delta\theta$*, and a uniform angular intensity distribution ( *i($\theta$) = 1.0* ), observed by a Young interferometer having a slit separation *a*. Borra(2008) shows that, for this simple model, the visibility of the spectral modulation at the beat frequency *$\omega$'* is given by a Bessel function having a maximum at *$\omega$' = 0* and its first zero given by the equation

$$\omega'a\Delta\theta/2c = 1.22 \, , \qquad (1)$$

where c is the speed of light. Therefore the spectral modulation of the output current has a maximum (highest visibility) at *$\omega$'=0* while the visibility disappears for values of *a* and *$\omega$'* given by Equation (1). Equation (1) illustrates how the interferometer works.

Because the beat frequency $\omega'$ and the slit separation *a* have the same effect, it shows that changing frequency $\omega'$ is equivalent to changing spacing *a* between the mirrors of an interferometer (like in amplitude or intensity interferometry); therefore, we can measure the angular diameter $\Delta\theta$ of the disk by increasing $\omega'$ until the visibility of the spectral modulation $I(\omega')$ of the output current disappears, while keeping *a* constant. It is very easy to vary $\omega'$, since $I(\omega')$ is directly obtained from the output current $I(t)$ of the quadratic detector. Figure 2 in Borra (2008) shows the decrease of the visibility as a function of ω' for the uniform disk model. The angular diameter $\Delta\theta$ is obtained from the value of $\omega'$ at which the visibility $V = 0$ (which is 1.22 for the uniform disk model) using Equation (1) above. For more complex cases, the angular intensity distribution can be obtained from the Fourier transform of the spatial coherence function $\gamma(a,\omega')$ which is obtained from the $\omega'$ dependence of the visibility function (Borra 2008).

This only gives a summary of the theoretical basis of the technique. To fully understand the technique one must read Borra (2008) that gives a complete theoretical justification based on wave-interactions. While a wave interaction theory, which is based on Maxwell equations, is fully justified, one must also take into account the fact that an electromagnetic wave has a granularity that comes from the quantization in photons. This adds photon noise to the signal. This issue is discussed in Borra (2008) and in section 3 that discusses application to Astronomical telescopes.

In the original experiment (Alford & Gold 1958) the spectral modulation was measured electronically with a short-wave receiver, while Basano & Ottonello (2000) used a wave analyzer and one could obviously use similar techniques. Modern technology however allows a more practical way to obtain the beat-spectrum that consists in first digitizing the output current $I(t)$ of the quadratic detector that measures the intensity signal from the interferometer and then performing with software the autocorrelation

$$I \otimes I = \int_{=\infty}^{\infty} I(t+t')I(t)dt = F(t') \quad . \tag{2}$$

The Wiener-Klintchine theorem then shows that the power spectrum can be obtained by taking the Fourier transform of the autocorrelation given by Equation (2). This procedure was used by Borra (2011) to obtain the power spectrum in the experiments that confirm the theoretical work in Borra (2008). Figure 3 in Borra (2011) shows what we mean by spectral modulation $I(\omega')$. The spectral modulation shown for a single optical path difference (OPD) (e.g. 3.12 –m) in Figure 3 in Borra (2011) is the kind of modulation one would see from a point-like source. While the spectral modulation shown for the sum of two OPDs is the kind of modulation one would see from a 2 point-like source, like a binary star. Note that the visibility of the spectral modulation does not vary noticeably with frequency in Figure 3 in Borra (2011) because the power spectrum displayed is at very low $\omega'/(2\pi)$ frequencies. The contrast would gradually decrease at much larger $\omega'/(2\pi)$ frequencies.

In the discussions so far, we have only considered the case where the optical beams from two separate telescopes are physically combined to optically interfere. I propose another interesting possibility that comes from separately digitally recording the output intensities $I_1(t)$ and $I_2(t)$ from the quadratic detectors at two distant telescopes and then sending the data to a central location where they are numerically added, giving $I(t) = I_1(t) + I_2(t)$, then performing the autocorrelation of $I(t)$ (Equation 2) and finally taking the Fourier transform of the autocorrelation. This procedure is justified below.

According to the Wiener-Klintchine theorem the power spectrum is obtained by taking the Fourier transform of the autocorrelation of a signal. In the autocorrelation of a time-dependent signal *I(t)*, all the information is obviously contained in the time dependence of *I(t)*. In particular the cosine spectral modulation generated by a point-like source (see equation 10 in Borra 2008), which is fundamental to us, comes from the fact that *I(t)* contains twin identical intensity fluctuations separated by a constant time $\tau$ given by the optical path difference of the interferometer. In the Alford &Gold (1958) experiment the signal comes from distinct pulses and the cosine modulation in the spectrum comes from the fact that the detector detects two identical intensity pulses separated by a time interval $\tau$. However, in the case of two continuous beams that are not pulsed and are superposed in an interferometer, the information is contained in intensity fluctuations due to wave interactions. The electric field *E(t)* and the intensity output of the quadratic detector *I(t)* fluctuate in time because of the superposition of electromagnetic waves that have different frequencies and varying phases. Klein & Furtak (1986) gives a convenient brief description of this effect. In an interferometer that uses a continuous source, the cosine modulation therefore comes from the fact that the interferometer splits the original intensity fluctuations in the input beam into twin identical fluctuations separated by $\tau$. The discussion in the introduction in Basano &Ottonello (2000), who repeated the Alford&Gold (1958) experiment using a continuous source, gives a heuristic justification for this procedure. As shown in the figure 2 in Borra (2011), that used a continuous source, the auto-correlation of the output current *I(t)* of a quadratic detector that measures the combined beams of two interferometers gives two identical peaked functions separated by twice the optical path difference (OPD) and another one at an OPD = 0. The relevant information is contained in the twin peaks with an OPD >0. Careful however that the figure 2 in Borra (2011), as mentioned in the figure caption, actually shows three separate autocorrelations identified by different colors.

Consider now that for a point-like astronomical source two separate telescopes will detect the same intensity fluctuations at two different times separated by

$$\tau_B = \boldsymbol{B}.\boldsymbol{s}/c, \qquad (3)$$

where $\boldsymbol{B}$ is the baseline vector (*B* is the distance between the telescopes and is therefore akin to *a* in Equation (1), *s* the unit vector in the direction of the source and *c* is the speed of light. If we numerically add the intensities $I_1(t)$ from a telescope and $I_2(t)$, from a second telescope, the same twin fluctuations separated by $\tau_B$ will be contained in $I_1(t) + I_2(t)$ as in the case where the optical beams are optically co-added in an interferometer. Using the digital output *I(t)* of optically combined interfering beams in Equation (2) is analogous to using the sum $I_1(t) + I_2(t)$ of the separately digitized $I_1(t)$ and $I_2(t)$ because the twin fluctuations separated by $\tau_B$, that carry the relevant information used in Equation (2) and its Fourier transform, have the same shapes and separations $\tau_B$ in both cases. This discussion is experimentally confirmed by Borra (2011) and clearly shown in his figures 2 and 3 (see caption of his figure 2).

As an important remark: Note that the Alford and Gold experiment did not use a classical interferometric set-up. The $E_1(t)$ and $E_2(t)$ signals were not coherent in the classical sense since, while they originated from the same spark, they were seen from two different mirrors that viewed the spark from different directions. For all practical purposes the $E_1(t)$ and $E_2(t)$ came from two separate sources. All of the relevant information was contained in the intensity signal shapes and the time delay $\tau_B$.

Another advantage of using the autocorrelation and the Fourier transform is that it is easy to filter out high frequency noise. This is done by restricting the limits of integration in the Fourier transform, a standard procedure in signal analysis.

### 3. APPLICATION TO ASTRONOMICAL TELESCOPES

A detailed discussion of applications to astronomical telescopes is beyond the scope of this article because it would have to dwell into technical details that differ greatly among telescopes that work at different frequencies of observation. Furthermore, technological advances over the next decades will certainly change the situation. We will instead limit ourselves to consider the advantages and disadvantages of the technique to astronomical applications in general and then consider, briefly, the application to two particular frequency regions: The optical and the radio regions.

a) Advantages

Its outstanding advantage is that it is far simpler than amplitude interferometry, as well as intensity interferometry, because the angular intensity distribution of a source is determined by measuring how the visibility of the spectral minima of the spectrum of intensity time fluctuations, measured with a quadratic detector, varies as a function of the beat-frequency $\omega'$. This is easy to do with software that first computes the autocorrelation of the intensity signal as a function of time and then computes the Fourier transform of the autocorrelation (see section 2). In amplitude and intensity interferometers, the angular intensity distribution of a source is obtained by changing the separation between individual mirrors (or using different telescopes at different locations). It obviously is far easier to change beat-frequency $\omega'$ with software that uses the technique based on the numerical autocorrelation of the intensity signal suggested in section 2, which uses the Fourier transform of the autocorrelation given by Equation (2), than physically move telescopes to change the baseline.

Note that, while changing the beat frequency is equivalent to changing the observational wavelength in an amplitude interferometer, there is a huge difference between the effectiveness of changing the beat frequency versus changing the observational wavelength. The observational wavelength can only be changed between the limits imposed by the detector (e.g. between 350 nm and 1000 nm for an optical telescope). This only allows a very small range of changes. On the other hand, in our case, the beat frequency can be changed to arbitrary low frequencies, well outside the bandpass of observation. For example, an optical telescope observing in the 350 to 1000 nm (35 to 100 THZ) region could easily get beat frequencies below 1 MHz. This can readily be seen in Borra (2011), where the detector observes at 1550 nm (65 THz) with a bandpass of 50 nm and power spectra are obtained at frequencies of a few hundred MHz.

One could also find the angular intensity distribution of a source without using Fourier transforms by carrying out numerical modeling of the cross-correlation signal. This would particularly be interesting for sources that have simple angular intensity distributions, like binary stars or the cores of many active nuclei of galaxies, quasars and BL Lac objects.

The technique should allow much higher angular resolutions than amplitude interferometry because, as discussed in section 2, the technique of separately digitally recording the output intensity signals at two distant telescopes and then sending the data to a central location where they are numerically added and then finally taking the Fourier transform of the autocorrelation would allow extremely large baselines. It could give extremely high angular resolutions, because a space interferometer could easily be used.

Another major advantage of the technique is that it allows for a more efficient use of telescope time because observations at a single baseline are sufficient since scanning is done by software. With standard interferometers one must change baseline and get many separate observations.

Having to move the telescopes sets a lower limit to the time variations that can be detected, since one obviously cannot detect variations below the time taken to move the telescopes. The fact that one does not have to move the telescopes, would therefore allow us to detect faster time variations.

Like with an intensity interferometer one can use inexpensive primary mirrors having surface qualities lower than those needed for amplitude interferometry. For the same reasons, the requirements for the alignment precision of all of the auxiliary optics are also less stringent than in amplitude interferometry. Consequently one could build dedicated inexpensive interferometers. The last paragraph in sub-section 3c that follows elaborates on this.

b) Disadvantages

Like for intensity interferometry, we consider wave interaction effects, hence one can obtain estimates of the signal to noise ratio by applying similar considerations. Photon shot-noise dominates when the counting rate is substantially below one count per coherence time interval $2\pi/\Delta\omega$. Like in intensity interferometry, it is therefore far easier to work in the radio region than in the optical since $2\pi/\Delta\omega$ is larger and the coherence times smaller in the optical. This issue is discussed at length in Borra (2008) and Borra (1997). This is the worst inconvenience of the technique in the optical-infrared region but *a minor inconvenience in the radio region.*

However, this limit is less severe than the hot bright star limit that applies to intensity interferometry (Hanbury Brown 1968). This can be understood from coherence theory, as discussed in p. 24 of Hanbury Brown (1968). In an intensity interferometer, when a star is unresolved, the signal-to-noise ratio increases with flux and therefore the area of the mirrors. However, as the diameters of the mirrors increase they become comparable with the baseline necessary to resolve the star, thereby reducing the coherence of the light at the detector and therefore $\gamma(0)$. Unfortunately, as the area increases, the increase in signal-to-noise ratio due to the increase in the total flux is counter-balanced by the loss in $\gamma(0)$ due to the decrease in coherence at the detector. Hanbury Brown (1968) calculates the maximum signal-to-noise ratio which can be obtained with two circular reflectors of unlimited size, on the assumption that the two mirrors are as close together as possible. The results, plotted in his figure 6, show that the maximum possible signal-to-noise ratio obtained with infinitely large reflectors is limited and varies with surface temperature making it difficult to observe objects having surface temperatures <4000° K. The new technique does not suffer from this problem for the fundamental reason that while the $\gamma(0)$ that applies to intensity interferometry depends on the frequency of observation $\omega$, and is therefore set by the detector, in our case $\gamma(0)$ varies with the beat-frequency $\omega'$. While $\omega$ remains constant as diameters and separation increase in intensity interferometry, in our case we are free to choose $\omega'$ and $\gamma(0)$ to suit our purpose. We can therefore work at arbitrarily lower $\omega'$ (and therefore higher $\gamma(0)$) with arbitrarily large mirrors.

   The limit is mitigated by the fact that, like in intensity interferometry (Hanbury Brown 1968) the requirements of the optical quality of the mirrors used are vastly more relaxed than for conventional interferometry so that poor quality but large reflectors can be used in the optical region.

   The disadvantage is also mitigated when one compares the present technique to amplitude interferometry in the optical region, because one needs a small bandpass in amplitude interferometry, while this limit does not apply to our technique, as discussed in Borra (2008). Furthermore, for amplitude interferometry in the optical region, a significant fraction of the light is lost in multiple reflections and refractions in auxiliary optics. In the technique discussed in section 2 that uses separately digitally recorded output intensity signals from two distant telescopes, there are no auxiliary optics.

   Note also that one could simultaneously observe in several different bandpasses, like in intensity interferometry, to minimize the problem. This comes about because , unlike in amplitude interferometry, the signal to noise ratio in wave interactions is independent of the bandpass (Hanbury Brown 1968) so that simultaneously observing in separate different small bandpasses increases the signal-to-noise ratio.

   A second major disadvantage comes from the fact that one measures at a lower frequency than the frequency of observation, one therefore needs a higher separation between telescopes to obtain the same angular resolution. Consequently, the telescopes must be connected over larger separations. This issue is discussed below for optical and radio-telescopes.

   c)        Infrared-Optical telescopes

Let us consider an interferometer that uses telescopes observing in the 1500 nanometer (200 THz) spectral region. The high beat-frequency limit is set by the speed limit (bandwidth) of the photodetector and electronics. Fortunately, there currently is massive effort to increase this speed limit because of important applications in telecommunications with fiber optics. A recent review of the technology is given by Beling & Campbell (2009). At the time the article was written, signals could be easily detected with a 0.2 THz bandwidth, with some signal detected at frequency as high as 0.4 THz. We can therefore assume that frequencies of the order of several THz should be measurable with future improvements.

For discussion purposes let us assume a detector having 0.5 THz frequency bandwidth. Using the Nyquist criterion we see that it could detect fluctuations at a frequency of 0.2 THz. Therefore an intensity fluctuation interferometer, observing at $v = \omega/2\pi = 200\ THz$. would need a baseline 1000 times larger to attain a resolution, at a beat-frequency $v' = \omega'/2\pi = 0.2\ THz$, equal to the resolution obtained at $v = 200\ THz$ with an amplitude interferometer . Consequently, to obtain a resolution comparable to the resolution attainable with an amplitude interferometer that uses two telescopes separated by 100 meters, comparable to the largest separation of existing optical interferometers, we would need two telescopes separated by 100 km. A 1000 km separation would give a factor of ten higher angular resolutions than presently achievable with existing optical interferometers. Telescopes separated by the Earth radius would have a resolution 65 times larger. Space interferometers would have even larger resolutions. Presumably, technological improvements in the bandwidth will allow much greater resolutions (and smaller baselines) in the future.

The quantity of data generated will however be very large and may cause problems since a properly sampled 0.2 THz bandwidth would generate of the order of 600 Gbits per second. This is a huge number but not discouragingly so. To appreciate this, consider the 2010 report in Photonics Spectra (Hogan 2011): Corning now commercially produces fiber optics capable of transporting 100 Gbs and 400 Gbs. It would not be a problem for a space interferometer that could send the data with light beams. One could also digitize the data, store it on an appropriate medium and transport it to a data center for later autocorrelation (see section 2).

When the 100 km distance required in fluctuation interferometry is compared to the 100-m distance required in optical amplitude interferometry for the same angular resolution, this may appear discouraging at first; however it is not, because of the other advantages that the technique brings in the optical-infrared region. For example, one could use a dedicated interferometer that uses inexpensive large Cherenkov telescopes. Dravins et al. (2012) discuss the science that can be done with an intensity interferometer that uses the Cherenkov Telescope Array. Lebohec & Holder (2006) estimates of the limiting magnitudes achievable with intensity interferometers and Cherenkov telescopes that also apply to the proposed interferometer. Table 1 in Lebohec & Holder (2006) shows that limiting magnitudes as faint as V = 7.8 can be reached with present Cherenkov telescopes and magnitudes as faint as V = 9.0 with the Next-Generation Cherenkov telescopes. Note that the end of the conclusion in Mandel (1962) clearly states that the signal to noise ratio in both intensity interferometers and the Alford and Gold effect, on which this interferometer is based (see section 2 above), is quantified by the degeneracy parameter. The degeneracy parameter quantifies the effect of photon shot noise in wave interaction physics, which applies to the proposed interferometer but most astronomers are unfamiliar with it. Borra (2008) gives a convenient brief summary of the quantification of the signal to noise ratio in terms of the degeneracy parameter.

The largest Cherenkov telescope (MAGIC) currently in operation has a 17-m diameter primary. Obviously having a dedicated interferometer with such large primary mirrors would have considerable interest. Another advantage is that one could get far more telescope time than what is available with the current amplitude interferometers, which are used with telescopes where the time is shared with other instruments. Consider also the time saving that occurs because a single observation at the single baseline is sufficient; while several observations with several separations are needed with amplitude interferometry.(see subsection 3a above). Finally future technological improvements in the bandwidth should allow to decrease the distance ratio.

One may worry about seeing effects in the optical region. Seeing effects are important in amplitude interferometry but not in our case. This can be understood by considering the discussion in Borra (2008) that shows that the relevant frequency is not the frequency of observation but the much lower beat-frequency. Note also that seeing effects are not important in intensity interferometry.

d) Radio-Telescopes

Let us now consider a radio telescope operating in the 0.1 cm to 10 cm (300 GHz to 3 GHz) spectral regions. Detectors capable of comparable bandwidths are commercially available. For example, the Pacific Millimeter Co makes detectors having 330 GHz bandwidth. Consequently one could operate with baselines that give angular resolutions comparable to those currently obtained with existing interferometers.

The proposed technique would particularly be useful if used with current VLBI networks and interferometers located in space. Consider that current VLBI techniques digitize the output electric field signal $E(t)$ from individual telescopes, digitally records it and then send it to a data processing correlator where the correlation is performed. The same data could be used to obtain the time-averaged intensity signal $I(t)$ from two telescopes needed for our purpose

$$I(t) = <[E_1(t) + E_2(t)]^2> \quad , \qquad (4)$$

where $E_1(t)$ and $E_2(t)$ are the electric fields measured at the separate telescopes and the brackets signify a time average. The intensity signal from Equation (4) would then have to be auto-correlated (Equation 2) and the Fourier transform obtained. Note in particular that this could be done with data obtained for other interferometric measures with current interferometers, including existing data. No new data would be needed. An interest of using the existing data is that one could detect time variations below the limit which was set by the time needed to move the telescopes, as discussed in subsection 3a.

With present radio telescopes, the principal advantage of the technique, as discussed in subsection 3a above, is that a single observation at a single baseline is sufficient since scanning is done by software, while with standard interferometers one must change baseline and get many separate observations. This makes a more efficient use of telescope time.

**4. CONCLUSION**

This article discusses a novel type of astronomical interferometer that uses measurements of the output intensity of a square-law detector (e.g. a photomultiplier) and has advantages over amplitude and intensity interferometers. An interesting suggestion is made in section 2: to separately record the intensity as a function of time at two different telescopes, then later numerically add them, and finally perform the autocorrelation of the added signals. This would allow having extremely large baselines and therefore extremely high angular resolutions.

The technique may be difficult to intuitively understand. However the discussion in section 2 leading to Equation (1) gives the basis for such an intuitive understanding. It considers, a simple source which is a circular disk having a uniform intensity distribution. Because the beat frequency $\omega'$ and the separation $a$ between the slits of a Young interferometer have the same effect, Equation (1) shows that changing frequency $\omega'$ is equivalent to changing the spacing $a$ between the mirrors of an interferometer, as is done in amplitude interferometry. Consequently, we can measure the diameter of the disk by increasing $\omega'$ until the spectral modulation at the beat frequency $\omega'$ disappears while keeping $a$ constant.

Simplicity is its outstanding advantage since the angular intensity distribution of an object is determined by measuring how the visibility of the spectral minima varies as a function of the beat-frequency $\omega'$ measured in the output current. This is easily done by first numerically performing the autocorrelation of the intensity measurements as a function of time and then taking the Fourier transform of the autocorrelation. In amplitude interferometers, the angular intensity distribution is determined by changing the separation between individual mirrors. As discussed in section 2, it is obviously far easier to change beat-frequency with software than physically move telescopes. This brings the major advantage of a more efficient use of telescope time because a single observation at a single baseline is sufficient, while amplitude interferometers need several observations at different baselines. Because one does not have to move the telescopes, the technique would also allow detecting faster time variations than classical interferometry.

Its major inconvenience in the optical region is that it needs very strong sources because it uses, like intensity interferometry, a wave-interaction effect. This is not a serious disadvantage in the radio region. This disadvantage is mitigated by the fact that, like in intensity interferometry, one could use inexpensive primary mirrors having surface qualities much lower than those needed for amplitude interferometry. Presently it could be useful in the optical region by applying it to observations with Cherenkov telescopes. This brings another major advantage in the optical-infrared region for one could use dedicated interferometers that use very large (tens of meters diameters) inexpensive telescopes.

In the radio region its present interest comes from the fact that it could improve the efficiency of existing telescopes because one could use measurements at a single baseline, while amplitude interferometry requires measurements at several different baselines.

Regarding the angular resolution that can be obtained with the technique; it cannot give a better resolution than amplitude interferometry at the same baseline separation of the telescopes. To the contrary, the resolution can only be worse since one measure at a lower beat frequency than the frequency of observation. However, in practice, it has an angular resolution advantage that comes from the fact that one could get data at considerably larger baselines than in amplitude interferometry. Therefore the baseline advantage would allow obtaining better angular resolution.

## ACKNOWLEDGEMENTS

This research has been supported by the Natural Sciences and Engineering Research Council of Canada.

### REFERENCES

Alford, W. P., & Gold, A., 1958, Am.J.Phys. , 26, 481
Basano, L., & Ottonello, P., 2000, Am.J.Phys., 68, 325
Beling, A. & Campbell, J.C. 2009, J. of Lightwave Technology 27, 343
Borra, E.F., 1997, MNRAS, 289, 660
Borra, E.F. 2008, MNRAS, 389, 364
Borra, E.F. 2011, MNRAS, 411, 1965
Dravins, D. et al. 2012, arXiv:1204.3624v1
Givens., M. P., 1961, J.Opt.Soc.Am., 51, 1030
Hogan, H., 2011, Photonics Spectra, 45, 63.
Hanbury Brown, R,.1968, Ann. Rev. Astr. Astrophys, 6,13
Klein, M. V., & Furtak, T. E., 1986, Optics (New York: John Wiley & Sons)
Lebohec, S. & Holder, J. 2006, ApJ 649, 399.
Mandel, L., 1962, J.Opt.Soc.Am., 52, 1335.